\documentclass[aps,prb,twocolumn,showpacs,groupedaddress]{revtex4}
\usepackage{graphicx}
\usepackage{natbib}
\usepackage[utf8]{inputenc}
\usepackage{epstopdf}
\setcitestyle{super}
\usepackage{color}
\usepackage[normalem]{ulem}

\begin{document}
\title{Disorder-enriched magnetic excitations in the Kitaev quantum spin liquid candidate Na$_2$Co$_2$TeO$_6$}
\author{Li Xiang}
\affiliation{National High Magnetic Field Laboratory, Tallahassee, Florida 32310, USA}
\author{Ramesh Dhakal}
\affiliation{Department of Physics and Center for Functional Materials, Wake Forest University, Winston-Salem, North Carolina 27109, USA}
\author{Mykhaylo Ozerov}
\affiliation{National High Magnetic Field Laboratory, Tallahassee, Florida 32310, USA}
\author{Yuxuan Jiang}
\affiliation{School of Physics and Optoelectronics, Anhui University, Hefei, Anhui 230601, China}
\affiliation{Center of Free Electron Laser and High Magnetic Field, Anhui University, Hefei 230601, China}
\author{Banasree S. Mou}
\affiliation{Department of Physics and Center for Functional Materials, Wake Forest University, Winston-Salem, North Carolina 27109, USA}
\author{Andrzej Ozarowski}
\affiliation{National High Magnetic Field Laboratory, Tallahassee, Florida 32310, USA}
\author{Qing Huang}
\affiliation{Department of Physics and Astronomy, University of Tennessee, Knoxville, Tennessee 37996, USA}
\author{Haidong Zhou}
\affiliation{Department of Physics and Astronomy, University of Tennessee, Knoxville, Tennessee 37996, USA}
\author{Jiyuan Fang}
\affiliation{School of Physics, Georgia Institute of Technology, Atlanta, Georgia 30332, USA}
\author{Stephen M. Winter}
\email{winters@wfu.edu}
\affiliation{Department of Physics and Center for Functional Materials, Wake Forest University, Winston-Salem, North Carolina 27109, USA}
\author{Zhigang Jiang}
\email{zhigang.jiang@physics.gatech.edu}
\affiliation{School of Physics, Georgia Institute of Technology, Atlanta, Georgia 30332, USA}
\author{Dmitry Smirnov}
\email{smirnov@magnet.fsu.edu}
\affiliation{National High Magnetic Field Laboratory, Tallahassee, Florida 32310, USA}
\date{\today}

\newcommand{\NCTO}{Na$_2$Co$_2$TeO$_6$ }

\begin{abstract}
Using optical magneto-spectroscopy, we investigate the magnetic excitations of Na$_2$Co$_2$TeO$_6$ 
in a broad magnetic field range ($0\ \rm{T}\leq B\leq 17.5\ \rm{T}$) at low temperature. Our measurements reveal rich spectra of in-plane magnetic excitations with a surprisingly large number of modes, even in the high-field spin-polarized state. Theoretical calculations find that the Na-occupation disorder in \NCTO plays a crucial role in generating these modes. Our work demonstrates the necessity to consider disorder in the spin environment in the search for Kitaev quantum spin liquid states in practicable materials.
\end{abstract}


\maketitle

Quantum spin liquids (QSLs) represent an intriguing phase of matter that can be formed by interacting quantum spins in certain magnetic materials
where quantum fluctuations impede the formation of long-range magnetic order even at the lowest temperatures \cite{Anderson1973,Kitaev2006,Balents2010a,Zhou2017,Broholm2020}. Kitaev's spin-1/2 honeycomb model \cite{Kitaev2006} has attracted considerable attention as it predicts an exotic QSL state with fractionalization of quantum spins into Majorana fermions in a potentially realizeable context for real materials. Among Kitaev candidate materials, $\alpha$-RuCl$_3$ is the most studied due to strong Kitaev interactions and signatures of a QSL state \cite{Banerjee2016,Do2017,Ran2017,Baek2017,Wang2017,Hentrich2018,Ponomaryov2020,Kasahara2018,Wulferding2020, Sahasrabudhe2020}. To describe the system, a generalized Heisenberg–Kitaev (gHK) model is frequently employed. 
Although the precise interactions remain debated \cite{Maksimov2020,laurell2020dynamical}, analysis of $\alpha$-RuCl$_3$ within the gHK model indicates significant departures from the ideal Kitaev model, prompting the search for new candidate materials.

Recently, with theoretical studies proposing that the $3d$ electrons with a high-spin $d^7$ configuration can provide pseudospin $J_{\rm eff}=1/2$ with Kitaev interactions \cite{Liu2018,Sano2018PRB,Liu2020PRL,doi:10.1142/S0217979221300061}, a Co-based honeycomb structure compound Na$_2$Co$_2$TeO$_6$
has been investigated extensively as another candidate to realize Kitaev physics \cite{Viciu2007,Lefrancois2016,Bera2017,Xiao2019,Hong2021}, together with other Co honeycomb compounds \cite{zhong2018field,nair2018short,zhong2020weak,das2021xy,kim2021spin,shi2021magnetic,maksimov2022ab,zhang2023magnetic,halloran2023geometrical}. Compared with $\alpha$-RuCl$_3$, \NCTO does not have observable stacking disorder between layers, but does have disorder in the interlayer Na positions \cite{Viciu2007,Xiao2019}. It exhibits a similar zigzag antiferromagnetic (AFM) order with $T_N \approx 27$ K. An in-plane magnetic field suppresses the magnetic ordering and was suggested to induce a QSL-like spin disordered state followed by a spin-polarized (SP) state above $B_c \approx 10$ T \cite{Hong2021,Lin2021}. However, recent elastic neutron scattering revealed that the zigzag Bragg peaks persist up to $B_c$ \cite{Yao2022arxiv}, thus questioning the intermediate QSL scenario. Early powder inelastic neutron scattering (INS) measurements were shown to be compatible with the gHK model, but reported parameters of competing Kitaev- and non-Kitaev interactions vary in a large range \cite{Songvilay2020PRB,Samarakoon2021,Kim2021,Sanders2022PRB}, the majority of which fail to reproduce the relative simplicity of the later-reported single-crystal INS data \cite{Chen2021,Sanders2022PRB,Yao2022}. These discrepancies have led to debate over the correct model, and even the magnetic ground state of \NCTO \cite{Chen2021,Yao2022arxiv,zhang2022electronic}. The robustness of Kitaev-dominant interactions has also been recently questioned \cite{kim2021spin,winter2022magnetic,liu2023non}.

In this Letter, we report on a systematic optical magneto-spectroscopy study of single-crystal \NCTO to probe the magnetic excitations in a broad magnetic field range at low temperatures. Our experiment reveals a rich set of in-plane excitation modes across the AFM to SP magnetic phases. Especially we observe a surprisingly large number of excitation modes in the SP state, which is in contrast to the case of $\alpha$-RuCl$_3$ and cannot be explained by any theoretical models proposed to date. Ab-initio calculations of the magnetic couplings and $g$-tensors together with polarized optical measurements reveal that the previously unconsidered Na-occupation disorder plays a crucial role in \NCTO. Theoretical results are further compared with literature-reported INS data and optical magneto-spectroscopy of this work to extract the interaction energies. With a comprehensive picture of the magnetic excitations as well as the non-trivial role of disorder, our work provides new insights and initiates re-thinking of the competing interactions in Kitaev QSL candidate materials.

\begin{figure*}[t]
\includegraphics[width=\linewidth]{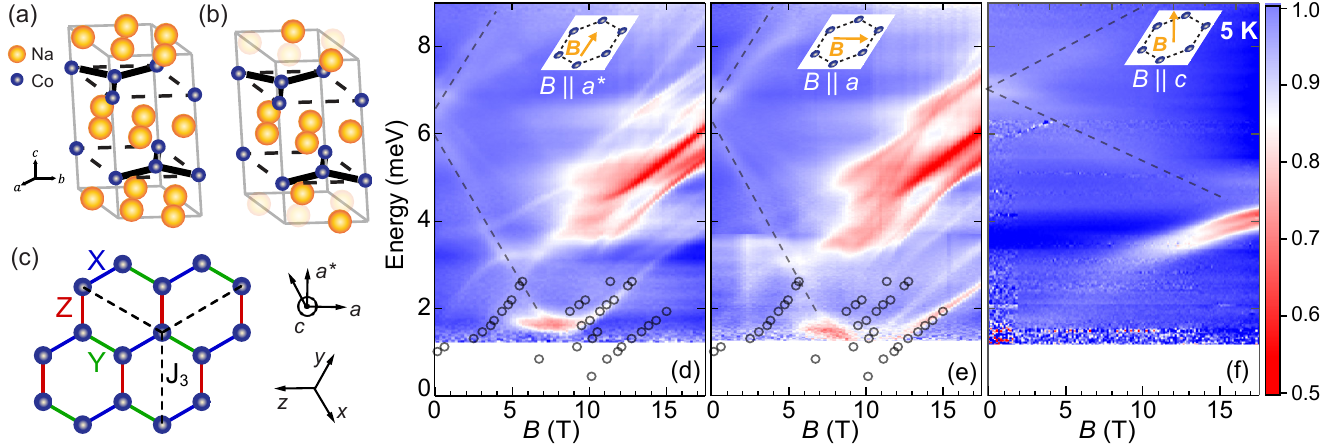}
\caption{(color online) \textbf{Unpolarized optical magneto-spectroscopy measurements on \NCTO.} (a) Average crystal structure with each Na site 2/3 occupied; Te and O atoms are omitted for clarity. (b) Representative structure showing Na-occupation disorder. See SM \cite{SM} for further details. (c) Honeycomb lattice of Co$^{2+}$ ions with nearest neighbor X, Y, Z and third neighbor $J_3$ bonds indicated. Crystallographic $(a,a^*,c)$ and cubic $(x,y,z)$ coordinates are shown. (d-f)
Normalized magneto-transmission measured on single-crystal \NCTO at $T = 5$ K with $B\parallel a^*$ (d), $B\parallel a$ (e), and $B\parallel c$ (f). Black circles are excitation modes extracted from the ESR spectra measured on powdered \NCTO samples at $T = 5$ K. Dashed lines are guides to the weak excitation modes at $\sim 6.5$ meV. The normalization procedure and raw ESR  spectra can be found in SM \cite{SM}.}
\label{fig1}
\end{figure*}

Millimeter-size \NCTO single crystals used in this study were grown by flux method \cite{Lin2021}. The crystal structure, following Ref. \cite{Xiao2019}, is illustrated in Fig.~\ref{fig1}(a). The reported $P6_322$ unit cell contains two formula units; each of the six Na sites is disordered with 2/3 occupied on average. A representative local realization of the Na disorder is shown in Fig.~\ref{fig1}(b). Far-infared (FIR) transmission spectroscopy was performed on single crystals. To probe the lower-energy excitations, electron spin resonance (ESR) spectroscopy was also carried out on powdered samples. More experimental details can be found in SM \cite{SM}.

Figures~\ref{fig1}(d-f) show the color map plots of the normalized FIR transmission intensity measured at $T=5$ K as a function of excitation energy and magnetic field applied $B\parallel a^*$, $B\parallel a$, and $B\parallel c$. The red color represents strong absorption. The excitation modes below 2.6 meV, extracted from the ESR spectra of powdered \NCTO samples, are plotted as black circles in Figs.~\ref{fig1}(d,e). The magnetic excitations revealed in ESR are well consistent with those observed in the FIR spectra measured on single-crystal samples with the magnetic field applied in-plane. Although the magneto-optical response of \NCTO summarized in Figs.~\ref{fig1}(d-f) displays a quite complex behavior, it is possible to identify three distinct spectral regions with respect to the suggested ($T$--$B$) phase diagram \cite{Lin2021,zhang2022electronic}: (I) Low-field region ($B<6$ T) with relatively weak excitations observed. Specifically, for in-plane magnetic fields, a zero-field gap of $\sim 1$ meV is revealed, which is attributed to magnons according to the reported INS data \cite{Yao2022}. In addition, excitations at $\sim 6.5$ meV, which split into two branches under applied magnetic field with different slopes between in- and out-of-plane (dashed lines in Figs.~\ref{fig1}(d-f)), are resolved for all field directions. (II) Intermediate-field region (6--9 T) with dispersionless modes at $\sim 1.7$ meV and in-plane field only, consistent with the proposed spin disordered phase region \cite{Lin2021}. (III) high-field ($B>10$ T)
SP state featuring strong magnon-like modes detected for all three field orientations. While two parallel excitation branches with slopes corresponding to $g_c\sim 3.7$ are detected for $B\parallel c$, the spectra of magnetic excitations with in-plane magnetic fields are much richer, indicating a strong anisotropy between in- and out-of-plane interactions. Furthermore, some subtle differences between $B\parallel a^*$ and $B\parallel a$ spectra at low-energy ($<$ 3 meV)  indicate the in-plane anisotropy as well. The temperature dependence of the FIR magneto-transmission up to 40 K is reported in SM \cite{SM}.

\begin{figure*}[t]
\includegraphics[width=\linewidth]{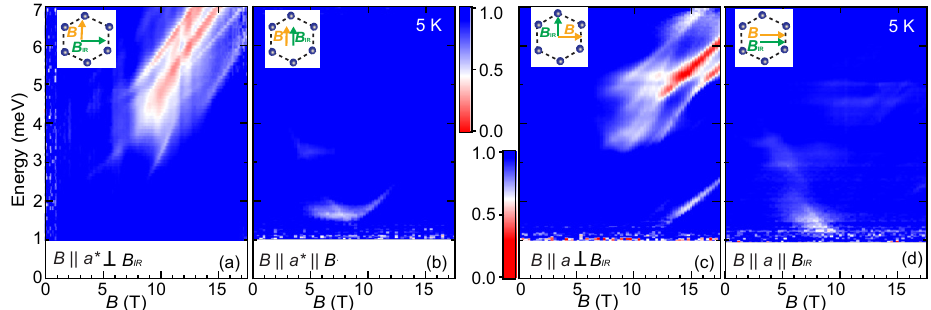}
\caption{(color online) \textbf{Polarized magneto-optical response of single-crystal \NCTO.} Color map plots show normalized FIR transmission intensity measured at $T = 5$ K 
and with $B \parallel a^*\perp B_{IR}$ (a), $B \parallel a^*\parallel B_{IR}$ (b), $B \parallel a\perp B_{IR}$ (c), and $B \parallel a\parallel B_{IR}$ (d). Top and bottom color bars are for plots in (a,b) and (c,d), respectively.}
\label{fig2}
\end{figure*}

The most striking feature of the FIR spectra is the richness of excitations in the high-field in-plane SP state, which are dominated by several strong absorption modes. This is in sharp contrast to the expectation for a pristine sample; the four Co atoms in the unit cell support a maximum of four magnon branches in the SP state. In practice, only one dominant absorption is expected, as two branches have weak intensity at $k=0$, and the other two would overlap energetically due to weak interplane coupling. Furthermore, the number of one-magnon modes would be expected to dramatically reduce below $B_c$, given that the low-field zigzag state is composed of multiple domains, each with a magnetic unit cell twice the volume of the crystallographic cell. These expectations are satisfied for $\alpha$-RuCl$_3$ \cite{Sahasrabudhe2020,Wulferding2020,Ponomaryov2020,Wu2018,Winter2018}, but not for \NCTO. 

In principle, the spectra of \NCTO may be enriched by multi-magnon and/or fractional excitations, as observed in $\alpha$-RuCl$_3$ \cite{Winter2017,Sahasrabudhe2020,Wulferding2020,Ponomaryov2020,Wu2018,Winter2018}. To investigate this possibility, polarized FIR spectroscopy measurements were carried out (Fig.~\ref{fig2}). Specifically, for both in-plane field directions, $B\parallel a^*$ (Figs.~\ref{fig2}(a,b)) and $B\parallel a$ (Figs.~\ref{fig2}(c,d)), the magnetic component of IR light $B_{IR}$ is polarized in the transverse $B\perp B_{IR}$ and longitudinal $B\parallel B_{IR}$ directions, which capture the dominant one-magnon and two-magnon channels, respectively, in the asymptotic high-field limit. The high-field in-plane magnetic excitations are pronounced only in the transverse $B\perp B_{IR}$ channel for both $B\parallel a^*$ and $B\parallel a$ directions. Moreover, all relatively strong excitations exhibit a similar d$E/$d$B$ slope with in-plane $g$-factor $g_{ab}\sim 4.6$. These results, in concert, suggest that the high-field in-plane magnetic excitations are one-magnon in nature. These findings are compatible with the reported magnetization data, which shows that \NCTO is nearly fully polarized at $B_c$ \cite{xiao2021magnetic,zhang2022electronic}; fluctuations that would be associated with strongly anisotropic couplings are weak, which calls into question the gHK model.

\begin{figure*}
\includegraphics[width=\linewidth]{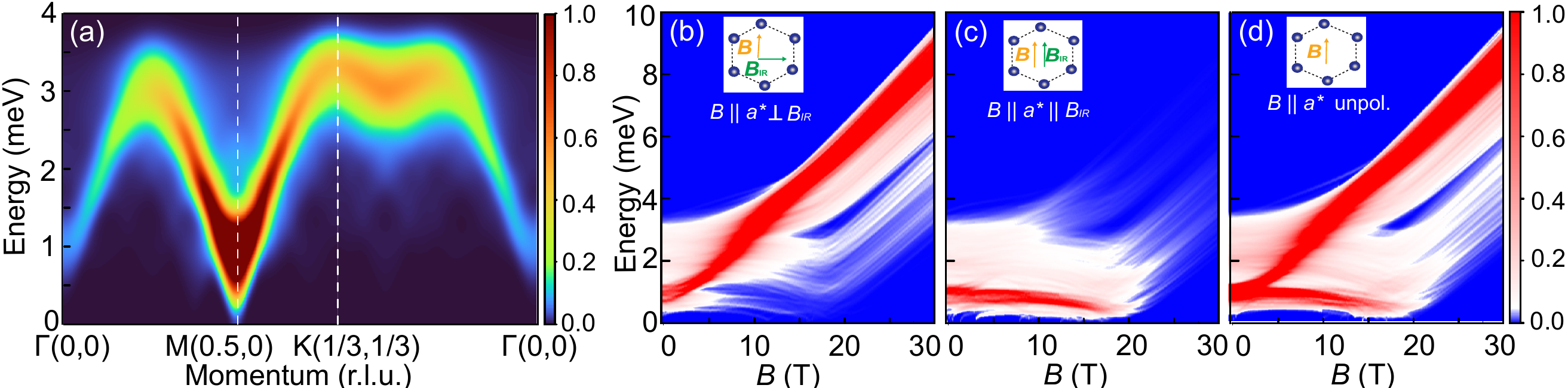}
\caption{(color online) \textbf{Model simulations of INS and FIR magneto-absorption spectra} for  random $J^1$ defined by the average values $\overline{J_x^1}, \overline{J_y^1} = -0.13$ meV, $\overline{J_z^1} = -0.24$ meV, $\overline{\Gamma_{xy}^1},\overline{\Gamma_{yx}^1} = -0.05$ meV, $\overline{\Gamma_{xz}^1},\overline{\Gamma_{zx}^1},\overline{\Gamma_{yz}^1},\overline{\Gamma_{zy}^1} = -0.08$ meV, and a range of values of width $0.8$ meV for each.  $J_3$ is Heisenberg matrix with $\overline{J_3}  = 1.8$ meV and a range of values of width $0.4$ meV. (a) Simulated INS spectra to be compared with Ref. \cite{Yao2022}. (b-d) Simulated optical magneto-spectroscopy spectra with the measurement geometry of $B (\parallel a^*)\perp B_{IR}$ (b), $B (\parallel a^*)\parallel B_{IR}$ (c), and $B \parallel a^*$ unpolarized (d). }
\label{fig3}
\end{figure*}

The experimental findings raise several questions: (i) what is the correct model for \NCTO, and (ii) what is the role of Na disorder? To address these, we investigated, with ab-initio calculations, the theoretical spread of magnetic couplings and $g$-tensors for different possible Na occupancies within the nominally $P6_3 22$ unit cell. Five unique Na distributions are identified, one of which is depicted in Fig.~\ref{fig1}(b) (see SM \cite{SM} for full details). For each structure, couplings were estimated by exact diagonalization of the full $d$-orbital model on two-site clusters and projection onto ideal $j_{1/2}$ states, according to the standard des Cloizeaux approach \cite{riedl2019ab,des1960extension}. This approach captures all bond-dependent anisotropic couplings \cite{winter2022magnetic}, but does not include additional ligand exchange processes. It is thus sufficient to demonstrate a strong effect of Na disorder on the couplings, but the ultimate model will be determined by fitting to experiment. Full calculation results are presented in \cite{SM}.

The $g$-tensors reflect the specific spin-orbital composition of the local moments, as determined by competing spin-orbit coupling (SOC) and crystal field splitting of the $t_{2g}$ orbitals of Co. At first consideration, one might expect disorder in the interlayer Na atoms to have little impact on the Co crystal fields. Despite the computed variation of the inter-$t_{2g}$ crystal field across different Co sites in the Na-disordered structures being only 2 to 18 meV, it is significant when compared to the small SOC strength in $3d$ Co, $\lambda_{\rm Co} \approx 60$ meV. As a consequence, the local $g$-tensors depend strongly on the random relative locations of the Na atoms around each Co; we find values in the range $g_a, g_b \sim 4.0$ to 5.8, and $g_c \sim 1.1$ to 2.9. The average computed in-plane value of $\overline{g_{ab}}\sim 4.9$ agrees well with the measured slope of the high-field excitations.

The magnetic couplings are $\mathcal{H} = \sum_{ij} \mathbf{S}_i \cdot \mathbf{J}_{ij} \cdot \mathbf{S}_j$, with 
$\mathbf{J}_Z = [ (J_x, \Gamma_{xy}, \Gamma_{xz}), (\Gamma_{yx}, J_y, \Gamma_{yz}), (\Gamma_{zx}, \Gamma_{zy}, J_z)]$ for the Z-bonds, where $J$ ($\Gamma$) are the on-diagonal (off-diagonal) exchange constants.
The global $(x,y,z)$ coordinate system is defined in Fig.~\ref{fig1}(c). The corresponding interactions for the X- and Y-bonds can be obtained by $C_3$ rotation along the $c$-axis. In the presence of Na disorder, the symmetry of each bond is reduced, such that it is convenient to describe the computed ranges of each coupling constant, denoted $R(...)$. For first neighbors, we estimate $R(J_x^1,J_y^1) = -0.6$ to 0.2 meV, $R(J_z^1) =$ 0.1 to 0.5 meV, $R(\Gamma_{xy}^1,\Gamma_{yx}^1) = $ 0.2 to 0.9 meV, and $R(\Gamma_{xz}^1,\Gamma_{zx}^1,\Gamma_{yz}^1,\Gamma_{zy}^1) = -0.3$ to 0.0 meV.  Thus, we find that the nearest neighbor interactions are considerably weaker than previous estimates from fitting the powder INS data \cite{Songvilay2020PRB,Samarakoon2021,Kim2021,Sanders2022PRB,Yao2022arxiv}, and moreover they vary with Na disorder on a scale similar to their overall magnitude. In contrast, for the third neighbor couplings, we find $R(J_\mu^3) = $ 2.3 to 3.5 meV and $R(\Gamma_{\mu\nu}^3) =-0.9$ to $-1.3$ meV. This corresponds to an AFM XXZ coupling. The employed method tends to overestimate the third neighbor couplings due to underlocalization of the DFT Wannier functions, but the finding of dominant third neighbor interactions is robust. See SM \cite{SM} for full results.

Before addressing the FIR/ESR spectra, given uncertainties in the model and structure of \NCTO, we first optimize the couplings to better fit reported single-crystal INS spectra \cite{Yao2022}. To this end, we consider a simpler model ignoring anisotropy in the third neighbor couplings. The nearest neighbor couplings are taken to be completely random for each bond (any spatial correlations associated with particular Na distributions are ignored), and are selected from a uniform distribution between prescribed limits. The $g$-tensors are also taken to be completely random symmetric tensors, with fixed ranges, in the global $(x,y,z)$ coordinates, following from the ab-initio results: $R(g_{\mu\nu}) = $ 3.0 to 4.8 for $\mu = \nu$ and $-1.3$ to $-0.6$ for $\mu \neq \nu$. The model is solved at the level of linear spin wave theory (LSWT) for $8\times 8 \times 1$ supercells and the resulting spectra are averaged over the three different zigzag domains and 20 different random choices of interactions. Results and details for the final model are shown in Fig.~\ref{fig3}. Despite having random interactions, the model has a zigzag ground state as a result of the dominant antiferromagnetic $J_3$ and nearest neighbor couplings that are ferromagnetic on average. The low-energy INS data is perfectly reproduced at the level of LSWT (Fig.~\ref{fig3}(a)). The smallness of the excitation gap at $k=0$ results from the overall weakness of the bond-dependent couplings, which is also compatible with the aforementioned immediate saturation of the magnetization at $B_c$ \cite{xiao2021magnetic,zhang2022electronic}. The existence of a single dominant magnon branch at zero field, despite overlapping response from three zigzag domains, can also be explained by the large $J_3$ model. With only $J_3$, the lattice is decoupled into three separate honeycomb sublattices. Different zigzag domains are related by global spin rotations on the decoupled sublattices, which cannot alter the magnon dispersions. Thus, in the $J_3$-only model, the zigzag domains have precisely identical magnon dispersions. A similar model was suggested in Ref. \cite{Yao2022}.

Finally, we address the polarized FIR spectra. LSWT results for the disordered model are shown in Figs.~\ref{fig3}(b-d). At zero field, the random couplings lead to a significant distribution of the excitation energies over the range of 1--4 meV, which is compatible with the experimental range. Since the disorder breaks all lattice symmetries, the number of observable modes is unchanged at $B_c$. As expected, the one-magnon modes rapidly lose intensity in the longitudinal channel ($B||B_{\rm IR}$) with increasing $B$ above $B_c$. While further elaboration of the model to include spatially correlated disorder may improve agreement with experiment in terms of the precise value of $B_c$ and number of distinct modes, it is already clear that the known Na disorder in \NCTO has a sufficient impact to explain the anomalously rich FIR/ESR spectra of Figs.~\ref{fig1} and \ref{fig2}. 

Our finding provides insight into several recent studies. In Ref. \cite{Yao2022}, the nearly dispersionless (i.e., local) excitations observed  at $\sim 6$ meV are difficult to reconcile with one-magnon excitations without fine tuning. Our FIR data (Figs.~\ref{fig1}(d-f) dashed lines) provides a possible identification. For all field orientations, this mode splits into two branches at finite $B$, with energies that follow roughly $dE/dB = \pm 2g$. For in-plane field, the lower branch appears to reach zero energy precisely at $B_c$. This is suggestive of local $\Delta S \approx \pm 2$ bound states in the zigzag phase, which are not captured by LSWT shown in Fig.~\ref{fig3}, but may gain intensity from disorder-induced non-collinearity of the spins. 

Regarding various complex \NCTO phase diagrams, a first-order transition is observed at low temperatures for $B\parallel a^*$ at $\sim 6$ T with significant hysteresis in the magnetization \cite{Xiao2019,Yao2020,xiao2021magnetic,Lin2021,lee2021multistage}. While initially interpreted as a possible QSL state \cite{Lin2021}, it is now revealed that well-defined zigzag Bragg peaks survive up to $B_c \approx 10$ T \cite{Yao2022arxiv}. Interestingly, the populations of the different zigzag domains in \NCTO are weakly field-dependent in comparison to $\alpha$-RuCl$_3$ \cite{sears2017phase,banerjee2018excitations}, which has contributed to speculation about an alternative single-domain triple-Q order \cite{Chen2021,lee2021multistage}. Our model favors zigzag order, as does an extensive recent experimental investigation \cite{zhang2022electronic}. The dominant interaction ($J_3$) does not have bond-orientation dependence, which leads to weaker energetic preference for moment orientation within a given zigzag domain, reducing the effectiveness of a field for selecting a particular domain. Furthermore, disordered couplings may locally select domains and pin domain walls, leading to hysteretic magnetization processes that do not constitute transitions out of zigzag order. 

 In conclusion, we have reported on the combined experimental magneto-IR/ESR spectroscopy and theoretical study of low-energy magnetic excitations in \NCTO. To explain the rich anisotropic spectrum of excitations observed in the experiment, we have developed a microscopic model that takes into account all unique possible Na occupancies. Complemented by the modeling of recent  INS data, we thus propose a comprehensive picture of magnetic excitations in \NCTO that reveals a key role of Na-occupation disorder. The findings highlight the fragility of spin-orbital moments in $3d$ compounds, where weak SOC enhances effects of disorder and distortions. In a more general context, our results further emphasize the necessity to consider disorder in the spin environment in the search for practicable materials potentially hosting Kitaev quantum spin liquid states.

\begin{acknowledgments} 
We thank Minseong Lee, Martin Mourigal, and Itamar Kimchi for insightful discussions. The experimental part of this work was primarily supported by the DOE (Grant No. DE-FG02-07ER46451), while the theoretical calculations were supported by the Center for Functional Materials, Wake Forest University. Computations were performed using the Wake Forest University (WFU) High Performance Computing Facility, a centrally managed computational resource available to WFU researchers including faculty, staff, students, and collaborators \cite{WakeHPC}. The crystal growth at UTK and the sample characterization at GT were both supported by the DOE (Grant Nos. DE-SC0020254 and DE-SC0023455, respectively). The IR measurements were performed at NHMFL, which is supported by the NSF Cooperative Agreement (Nos. DMR-1644779 and DMR-2128556) and the State of Florida. Y.J. acknowledges support from the National Natural Science Foundation of China (Grant No. 12274001) and the Nature Science Foundation of Anhui Province (Grant No. 2208085MA09).

L.X. and R.D. contributed equally to this work.
\end{acknowledgments}

\bibliographystyle{apsrev}

\end{document}